# Unifying Magnetic, Electrical, and Thermoelectric Responses in a Non-Collinear Antiferromagnet

André José[1,2], Adrielson Dias[1,2], Carlos Eduardo[2,3], Giovanna Celly[2,3], Luiza Paffer[1,2], José Araújo[2,4], José Laurentino[1,2], Monalisa Cavalcante[1,2], and José Holanda[1,2,3,4,5,*]

[1]Programa de Pós-Graduação em Engenharia Física, Universidade Federal Rural de Pernambuco, 54518-430, Cabo de Santo Agostinho, Pernambuco, Brazil
[2]Group of Optoelectronics and Spintronics, Universidade Federal Rural de Pernambuco, 54518-430, Cabo de Santo Agostinho, Pernambuco, Brazil
[3]Programa de Pós-Graduação em Física Aplicada, Universidade Federal Rural de Pernambuco, 52171-900, Recife, Pernambuco, Brazil
[4]Unidade Acadêmica do Cabo de Santo Agostinho, Universidade Federal Rural de Pernambuco, 54518-430, Cabo de Santo Agostinho, Pernambuco, Brazil
[5]Programa de Pós Graduação em Tecnologias Energéticas e Nucleares (Proten), Universidade Federal de Pernambuco, Recife, 50740-545, PE, Brazil

## Abstract

Non-collinear antiferromagnets offer unconventional routes for controlling magnetic, electrical, and thermoelectric responses through their complex spin configurations and interfacial symmetry. Here, we detect exceptional interfacial properties arising from a non-collinear antiferromagnet by demonstrating a robust and quantitatively consistent exchange-bias response in an $IrMn_3$/Py heterostructure. The system consists of a 10 nm $IrMn_3$ layer coupled to a 5 nm permalloy (Py) film. Longitudinal magneto-optical Kerr effect (MOKE), anisotropic magnetoresistance (AMR), and anomalous Nernst effect (ANE) measurements independently reveal the same unidirectional field shift of approximately 50 Oe, together with a consistent cosine angular dependence. The simultaneous observation of these signatures establishes a direct correspondence between the interfacial magnetic symmetry imposed by the non-collinear antiferromagnet and the responses of the adjacent ferromagnet. This symmetry is manifested consistently in magnetization reversal, charge transport, and thermally driven voltage generation, demonstrating that the exchange-bias imprint is not restricted to a single experimental observable. Our results reveal a coherent multifunctional response of $IrMn_3$/Py and extend the conventional understanding of exchange bias beyond collinear antiferromagnetic systems. More broadly, they demonstrate the potential of non-collinear antiferromagnets as platforms for coupling magnetic, electrical, and thermoelectric functionalities in spintronic and spin-caloritronic devices.

*Corresponding author: joseholanda.silvajunior@ufrpe.br

Orcid: https://orcid.org/0000-0002-8823-368X

## 1. Introduction

Antiferromagnetic materials have become a major focus of modern spintronics because they offer a combination of properties that are difficult to achieve in conventional ferromagnets: negligible stray magnetic fields, robustness against external magnetic perturbations, ultrafast spin dynamics, and high integration density [1–4]. In a compensated antiferromagnetic state, the magnetic moments of different sublattices cancel macroscopically, eliminating dipolar cross-talk between neighboring devices while preserving long-range magnetic order. These characteristics make antiferromagnets highly attractive for high-speed memory, logic, sensing, and nanoscale information technologies. In conventional collinear antiferromagnets, neighboring moments are aligned antiparallel along a common axis, and their physical behavior is often described using two oppositely oriented magnetic sublattices [3, 4]. Such systems have historically played a key technological role through the exchange-bias effect, in which coupling to a ferromagnetic layer shifts the magnetic hysteresis loop and establishes a preferred direction for magnetization reversal [5–14]. This phenomenon has been extensively exploited in spin valves, magnetic tunnel junctions, and read-head technologies [15-20].

More recently, attention has shifted toward non-collinear antiferromagnets, in which the local magnetic moments are not restricted to antiparallel alignment but instead form angular spin textures such as triangular, kagome, chiral, or frustrated arrangements [21-27]. In these systems, the vector sum of the magnetic moments may remain zero, while the spin configuration breaks time-reversal and crystalline symmetries in unconventional ways [28-36]. As a consequence, non-collinear antiferromagnets can host functional transport phenomena that were once thought to require finite magnetization, including anomalous Hall effects, spin Hall conversion, current-induced torques, and anomalous Nernst responses [4, 15–18, 30, 36–41]. This realization has transformed the role of antiferromagnets in spintronics. Rather than acting solely as passive pinning layers, non-collinear antiferromagnets can actively generate, manipulate, and detect spin and charge currents. Their compensated magnetic state combined with symmetry-enabled transport effects offers a powerful route toward low-power multifunctional devices.

Among this class of materials, $IrMn_3$ has emerged as a prototypical metallic non-collinear antiferromagnet of both fundamental and technological importance. In $IrMn_3$, the Mn moments form a triangular spin structure stabilized by exchange interactions and spin-orbit coupling, producing a compensated magnetic ground state with strong anisotropic and relativistic transport signatures [4, 42]. This magnetic structure is fundamentally distinct from that of collinear exchange-bias materials such as FeMn or NiO. Because of its triangular spin texture, $IrMn_3$ can generate symmetry-selective spin currents and large spin Hall effects, while simultaneously preserving the practical advantages of zero net magnetization. IrMn-based alloys are already widely employed in industrial spintronic architectures as exchange-bias layers. However, many previous studies have treated these materials within the conventional framework of static ferromagnetic pinning, without explicitly considering the consequences of their non-collinear magnetic order. This leaves open an important question: to what extent does the triangular spin texture of $IrMn_3$ influence the magnitude, angular dependence, and multifunctional signatures of exchange bias at a ferromagnetic interface?

At a ferromagnet/non-collinear antiferromagnet interface, the exchange field may arise not only from uncompensated interfacial moments, but also from spin canting, directional spin projections, local symmetry breaking, and spin-orbit-mediated anisotropies inherited from the antiferromagnetic texture. Consequently, exchange bias in these systems may be simultaneously encoded in magnetic, electrical, and thermoelectric observables, extending far beyond the traditional picture of a shifted hysteresis loop. The magnetic signature of exchange bias is classically obtained from the displacement of the hysteresis loop by a field $H_{EB}$, defined as $H_{EB} = (H_{C1} + H_{C2})/2$, where $H_{C1}$ and $H_{C2}$ are the coercive fields for decreasing and increasing magnetic field sweeps, respectively [5–7]. In a strongly anisotropic interface, $H_{EB}$ is expected to depend on the projection of the unidirectional anisotropy axis onto the applied-field direction, commonly following $H_{EB}(\phi_H) = H_0 \cos\phi_H$, where $\phi_H$ is the in-plane field angle and $H_0$ is the maximum exchange-bias amplitude.

Electrical access to this interfacial state can be obtained through anisotropic magnetoresistance (AMR), where the resistance depends on the angle between electric current and magnetization according to [15, 16] $R(\phi)=R_\perp+\Delta R_{AMR}\cos^2\phi$, with $R_\perp$ the resistance for magnetization perpendicular to the current and $\Delta R_{AMR} = R_\parallel - R_\perp$ the AMR amplitude. Since exchange bias shifts the magnetization reversal field, the same interfacial anisotropy should be detectable in AMR loops. A complementary thermoelectric probe is provided by the anomalous Nernst effect (ANE), in which an applied temperature gradient generates a transverse electric field given by [33–46] $\mathbf{E}_{ANE} = \alpha_N(\nabla T\times\mathbf{m})$, where $\alpha_N$ is the anomalous Nernst coefficient and $\mathbf{m}$ is the unit vector along the magnetization. Because the ANE voltage depends directly on magnetic orientation, it offers a sensitive route to detect exchange-pinned states under nonequilibrium thermal conditions.

Although MOKE, AMR, and ANE have each been extensively studied in magnetic thin films, their combined use to probe exchange bias imposed by a non-collinear antiferromagnet remains largely unexplored. Demonstrating quantitative agreement among these three techniques would provide compelling evidence that the same interfacial symmetry-breaking field governs magnetic reversal, charge transport, and thermally driven voltage generation. In this work, we investigate exchange coupling in $IrMn_3$/Py heterostructures composed of a 10 nm $IrMn_3$ layer and a 5 nm permalloy (Py) film. Using longitudinal magneto-optical Kerr effect (MOKE), anisotropic magnetoresistance (AMR), and anomalous Nernst effect (ANE) measurements, we demonstrate that the exchange-bias field is consistently measurable through all three probes, with identical magnitude and angular dependence. Our results establish that $IrMn_3$ is not merely a conventional exchange-bias material, but an active non-collinear antiferromagnet whose triangular spin symmetry is transferred into magnetic, electrical, and thermoelectric responses. These findings broaden the conventional concept of exchange bias and position $IrMn_3$-based heterostructures as promising platforms for next-generation spintronic, spin-caloritronic, and multifunctional magnetic technologies [22, 23, 29, 31].

## 2. Experimental Methods

The investigated heterostructures were fabricated by sequential deposition of a 10 nm-thick $IrMn_3$ layer followed by a 5 nm-thick permalloy (Py, $Ni_{80}Fe_{20}$) film using dc magnetron sputtering under an Ar pressure of 2 mTorr and a base pressure of $10^{-9}$ Torr. The films had lateral dimensions of 4×2 $mm^2$. Particular attention was given to the growth of the $IrMn_3$ layer, since this material is a prototypical non-collinear antiferromagnet whose magnetic behavior originates from a triangular arrangement of Mn moments. In this spin configuration, the local magnetic moments remain nearly compensated macroscopically, while preserving strong internal exchange fields and pronounced symmetry-dependent anisotropies [4, 42]. During deposition, an in-plane magnetic field of 0.5 kOe was applied along the [100] crystallographic direction. This procedure was used to define the exchange-bias axis and to promote preferential alignment of interfacial spin projections within the (001) plane of $IrMn_3$. Although $IrMn_3$ exhibits no net bulk magnetization, its non-collinear magnetic texture can generate symmetry-selective uncompensated or weakly compensated spin components at the interface. These interfacial moments are expected to couple efficiently to the adjacent Py layer, producing a stable unidirectional anisotropy after growth. Magnetization measurements were performed by longitudinal magneto-optical Kerr effect (MOKE), with the external magnetic field applied in the plane of the sample. Hysteresis loops M(H) were recorded as a function of the in-plane field angle $\phi_H$, measured relative to the [100] direction, as illustrated in **Fig. 1**. The exchange-bias field was extracted from the loop displacement according to $H_{EB} = (H_{C1}+H_{C2})/2$, where $H_{C1}$ and $H_{C2}$ are the coercive fields for decreasing and increasing magnetic-field sweeps, respectively. The angular evolution of $H_{EB}$ provides direct information on how the anisotropic spin texture of $IrMn_3$ is projected onto the ferromagnetic reversal process. Anisotropic magnetoresistance (AMR) measurements were carried out using a dc current of 0.1 mA in a pseudo-four-probe geometry. This configuration allowed the same contacts to be used in both transport and thermoelectric measurements, minimizing systematic errors associated with contact resistance or geometric asymmetry. Because AMR depends on the angle between current and magnetization, it serves as an electrical probe of the Py magnetization trajectory under the influence of the exchange field generated by the non-collinear $IrMn_3$ layer. The anomalous Nernst effect (ANE) was measured by imposing an out-of-plane temperature gradient through a Peltier element positioned beneath the sample. Under these conditions, a transverse thermoelectric voltage is generated in the Py layer according to $\mathbf{E_{ANE}} = \alpha_N (\mathbf{\nabla} T\times\mathbf{m})$, where $\alpha_N$ is the anomalous Nernst coefficient and $\mathbf{m}$ is the unit vector along the magnetization direction [33–38]. The ANE voltage was measured as a function of magnetic-field amplitude and in-plane angle $\phi_H$, allowing the exchange-bias field to be independently determined from the shift of the thermovoltage loops. Additional ANE measurements were performed under fixed temperature gradients while sweeping the magnetic field above magnetic saturation, enabling extraction of $\alpha_N$. Since ANE directly couples heat flow to magnetic order, it provides complementary access to how the interfacial anisotropy imposed by $IrMn_3$ remains active under nonequilibrium thermal conditions. The combination of MOKE, AMR, and ANE is particularly powerful for the present system because it probes the same interfacial exchange state through three distinct channels: magnetic reversal, charge transport, and thermoelectric conversion. This multiprobe methodology is especially relevant for non-collinear antiferromagnets such as $IrMn_3$, whose complex spin symmetry can manifest simultaneously in magnetic and transport observables.

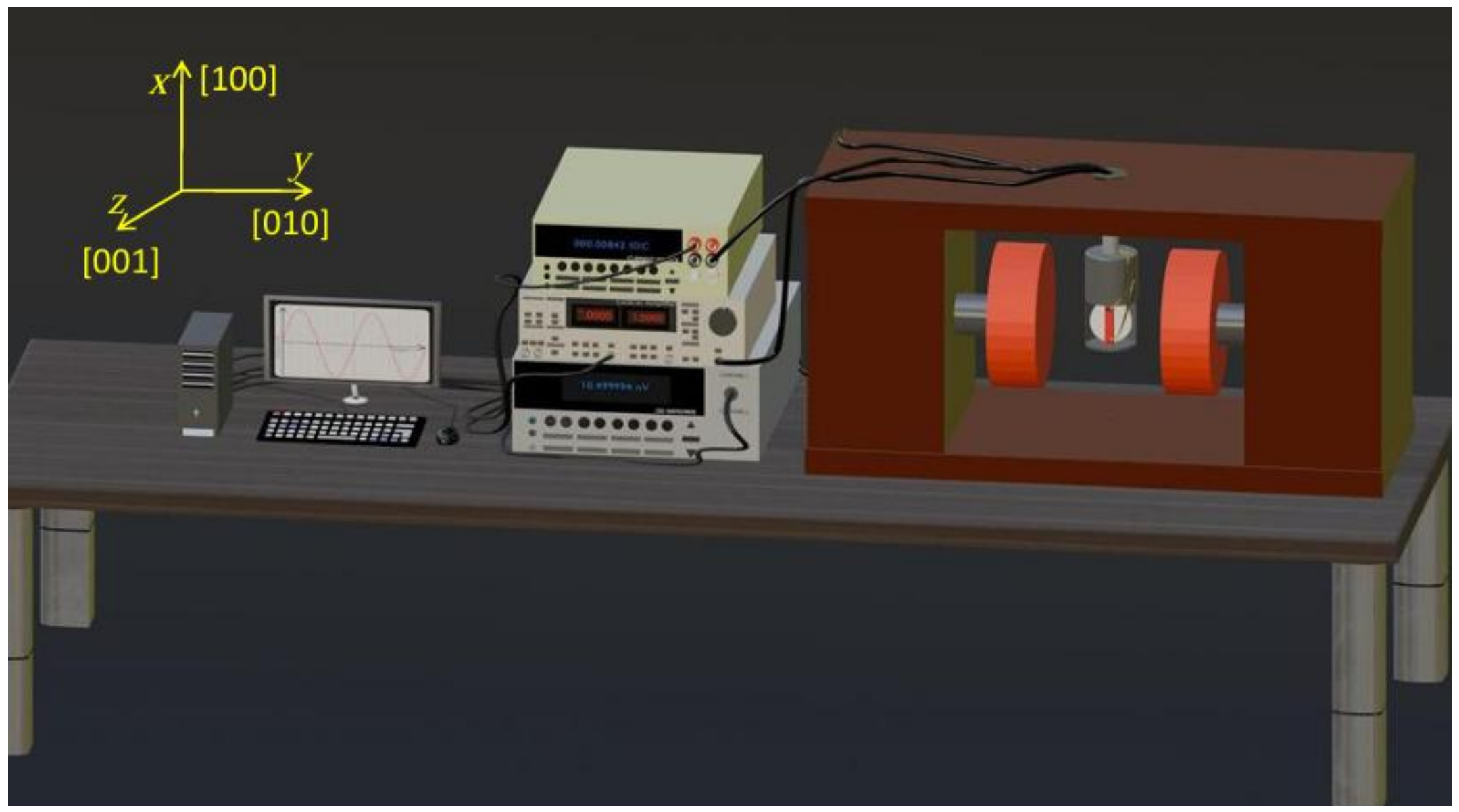

**Figure 1**: Experimental configuration employed for longitudinal magneto-optical Kerr effect (MOKE), anisotropic magnetoresistance (AMR), and anomalous Nernst effect (ANE) measurements in the $IrMn_3$/Py heterostructure. The crystallographic coordinate system is defined by the in-plane directions [100] and [010], and the out-of-plane direction [001]. This symmetry framework determines the relative orientations of the applied magnetic field, electric current, and temperature gradient, and is essential for resolving the anisotropic magnetic response and exchange-bias effects imposed by the non-collinear antiferromagnetic $IrMn_3$ layer.

## 3. Results and Discussion

The magnetic, magnetotransport, and thermoelectric responses of the $IrMn_3$/Py bilayer were investigated by combining longitudinal magneto-optical Kerr effect (MOKE), anisotropic magnetoresistance (AMR), and anomalous Nernst effect (ANE) measurements. This multi-probe strategy is particularly powerful for $IrMn_3$-based heterostructures because $IrMn_3$ is not a conventional passive exchange-bias material, but a metallic non-collinear antiferromagnet whose triangular Mn spin structure can imprint anisotropic interfacial exchange fields while also exhibiting strong spin-dependent transport properties [4, 42]. The combined results show that exchange bias in the present system is consistently encoded in magnetic reversal, electrical resistance, and thermally generated voltage, establishing a unified picture of interfacial symmetry breaking. Longitudinal MOKE measurements were first performed with the magnetic field applied in the plane of the sample. The exchange-bias field $H_{EB}$ was extracted from the horizontal displacement of the M(H) hysteresis loops as a function of the angle $\phi_H$ between the applied field and the easy axis. For $\phi_H$=0°, aligned with the crystallographic [100] direction [**Fig. 2(a)**], the hysteresis loop exhibits a clear horizontal shift corresponding to $H_{EB}$=50 Oe. The square loop shape and well-defined coercive fields indicate a coherent reversal process stabilized by a robust unidirectional anisotropy imposed by the $IrMn_3$ layer.

This result is significant because exchange-bias fields in the range of a few tens of Oe are commonly reported for soft ferromagnets exchange-coupled to thin IrMn-based antiferromagnetic layers, depending strongly on thickness, texture, growth field, and interfacial roughness [5–14]. Therefore, the present value of 50 Oe confirms that the non-collinear magnetic order of $IrMn_3$ is fully capable of generating technologically relevant pinning fields comparable to conventional exchange-bias systems. Rather than suppressing interfacial coupling, the triangular spin texture of $IrMn_3$ appears to sustain efficient exchange transfer to the Py layer. For $\phi_H$=90∘ [**Fig. 2(b)**], the hysteresis loop becomes nearly symmetric with $H_{EB}$≈0, identifying the hard-axis orientation. In this geometry, the projection of the unidirectional exchange field onto the external field direction vanishes, as expected for a well-defined exchange axis. The pronounced angular contrast between 0∘ and 90∘ demonstrates that the bias field is highly directional and not associated with trivial remanent offsets or isotropic defects. Instead, it reflects a crystallographically correlated interfacial anisotropy originating from the $IrMn_3$ spin structure. These observations are particularly relevant in the context of non-collinear antiferromagnets. Because the Mn moments in $IrMn_3$ form a triangular arrangement, the interface can host uncompensated or weakly compensated spin projections even when the bulk moment remains zero [4, 42]. Such symmetry-selective interfacial moments provide a natural microscopic mechanism for the observed deterministic exchange bias.

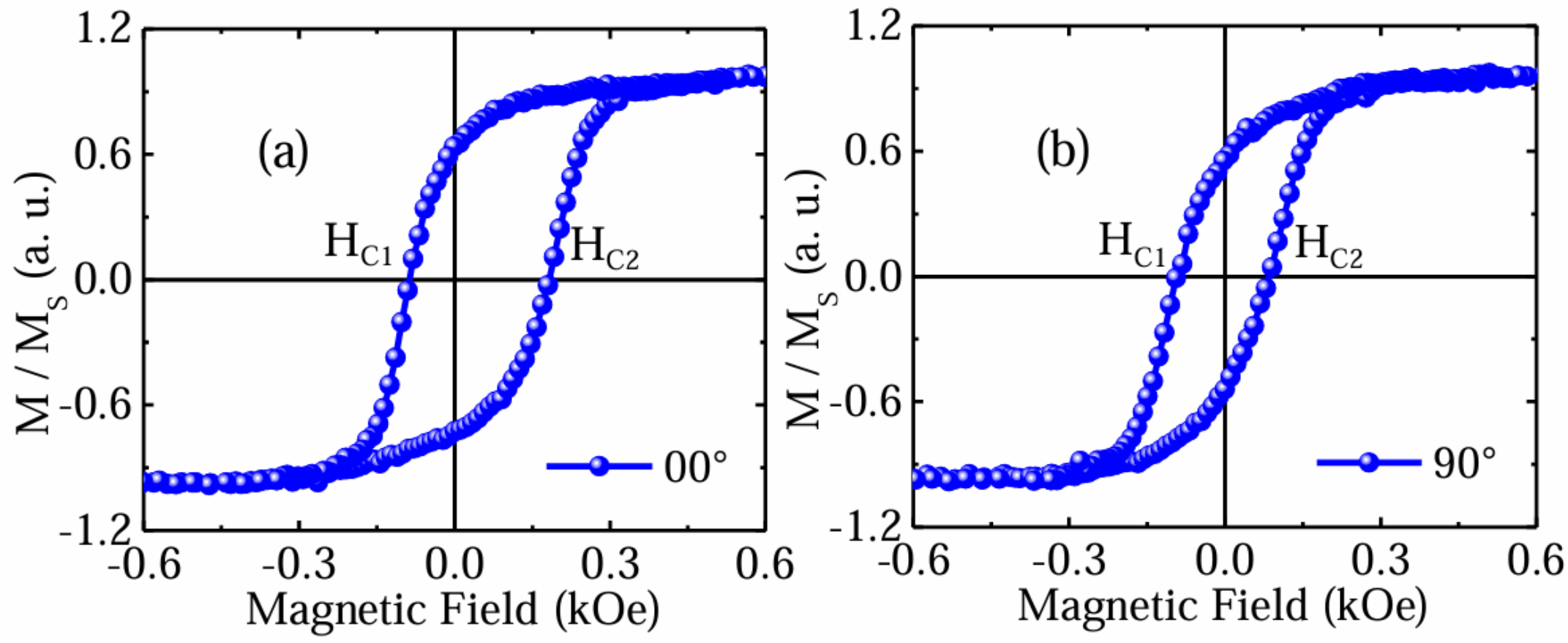

**Figure 2**: Longitudinal MOKE hysteresis loops measured with the magnetic field applied in-plane for **(a)** $\phi_H$=0∘ (easy axis), and **(b)** $\phi_H$=90∘ (hard axis), evidencing the angular dependence of the exchange-bias field.

The exchange-bias field can also be resolved electrically through anisotropic magnetoresistance measurements. A dc current of 0.1 mA was applied using a pseudo-four-probe geometry, with identical contacts employed for both current injection and voltage detection. This configuration ensures the same electrical pathway for subsequent ANE measurements and minimizes geometrical artifacts. The electrode spacing was 2 mm. **Fig. 3(a)** displays the AMR curves for $\phi_H$=0∘ and $\phi_H$=90∘. In both cases, the resistance traces exhibit a clear field displacement consistent with the exchange bias independently obtained from MOKE. This agreement demonstrates that the interfacial exchange field imposed by $IrMn_3$ not only shifts the magnetic hysteresis loop but also governs the transport state of the Py layer. This correspondence between MOKE and AMR is nontrivial. MOKE directly probes magnetization reversal, whereas AMR measures the angular dependence of spin-dependent scattering. The fact that both techniques yield the same bias field indicates that the exchange coupling rigidly controls the magnetization trajectory responsible for charge transport. Such electrically readable exchange bias is highly attractive for integrated devices, where resistive detection is more practical than optical probing.

The angular dependence of the resistance at a fixed magnetic field of 0.9 kOe is shown in **Fig. 3(b)**. The data follow the standard AMR relation $R(\phi)=R_{\perp}+\Delta R_{AMR}\cos^2\phi_H$, where $R_{\perp}$ is the resistance with magnetization perpendicular to the current and $\Delta R_{AMR}=R_{\parallel}-R_{\perp}$ is the AMR amplitude. The fitting yields $R_{\perp}$=86.2 Ω and $\Delta R_{AMR}$=−53 mΩ. The magnitude of the AMR ratio is consistent with typical values reported for thin Py layers, indicating that exchange coupling to $IrMn_3$ does not degrade the intrinsic transport properties of the ferromagnet. Instead, $IrMn_3$ acts as a functional antiferromagnetic underlayer that stabilizes the magnetic state while preserving conventional metallic magnetotransport.

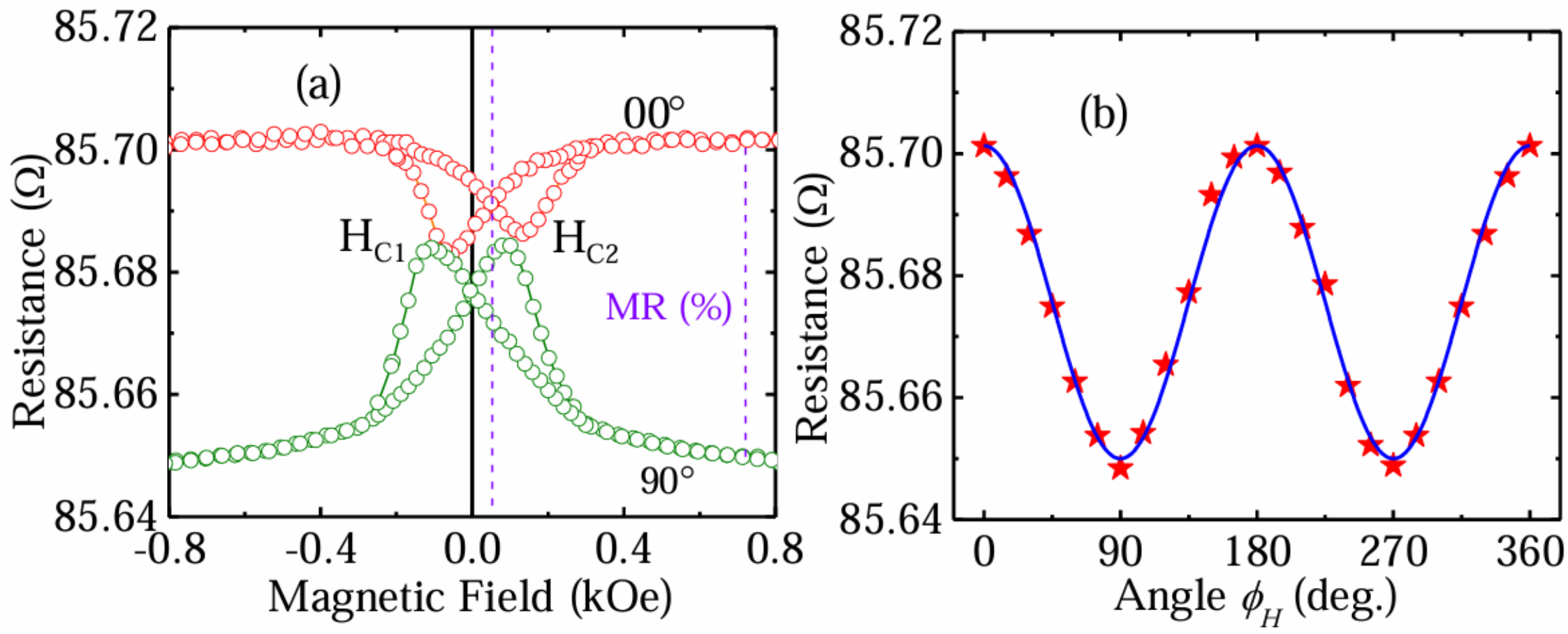

**Figure 3**: **(a)** Anisotropic magnetoresistance (AMR) measured with the magnetic field applied at $\phi_H$=0° and $\phi_H$=90∘, highlighting the field-shifted resistance traces associated with exchange bias. **(b)** Angular dependence of the AMR at H=0.9 kOe, exhibiting the characteristic $\cos^2\phi_H$ symmetry.

The anomalous Nernst effect was then used as a thermoelectric probe of the same interfacial exchange state. An out-of-plane temperature gradient was applied across the bilayer using a Peltier element, while the electrical configuration remained identical to that used for AMR measurements. **Fig. 4(a)** presents the ANE voltage as a function of magnetic field for different angular orientations under a fixed temperature gradient $\Delta T_z$ =10 K. The voltage reverses sign upon magnetization switching, confirming its anomalous spin-dependent origin. Most importantly, the ANE loops also exhibit a clear horizontal displacement corresponding to the same exchange-bias field observed in MOKE and AMR. This result demonstrates that the exchange-pinned state generated by $IrMn_3$ remains fully active under nonequilibrium thermal conditions. In other words, the non-collinear antiferromagnetic layer controls not only magnetic reversal and electrical transport, but also thermally driven transverse voltage generation.

The angular dependence measured at H=0.9 kOe [**Fig. 4(b)**] follows $V_{ANE}(\phi_H)=V_{0y}^{ANE}+V_y^{ANE}\cos\phi_H$, consistent with $\mathbf{E_{ANE}}=\alpha_N(\nabla T_z\times\mathbf{m})$. The fit yields a negligible offset voltage $V_{0y}^{ANE}\approx0$, indicating minimal parasitic thermoelectric contributions, and an angular modulation amplitude $V_y^{ANE}=0.9$ μV. Such clean cosine symmetry confirms that the measured signal is dominated by the intrinsic anomalous Nernst response of the Py layer, while the $IrMn_3$ interface defines the magnetic switching field through exchange bias.

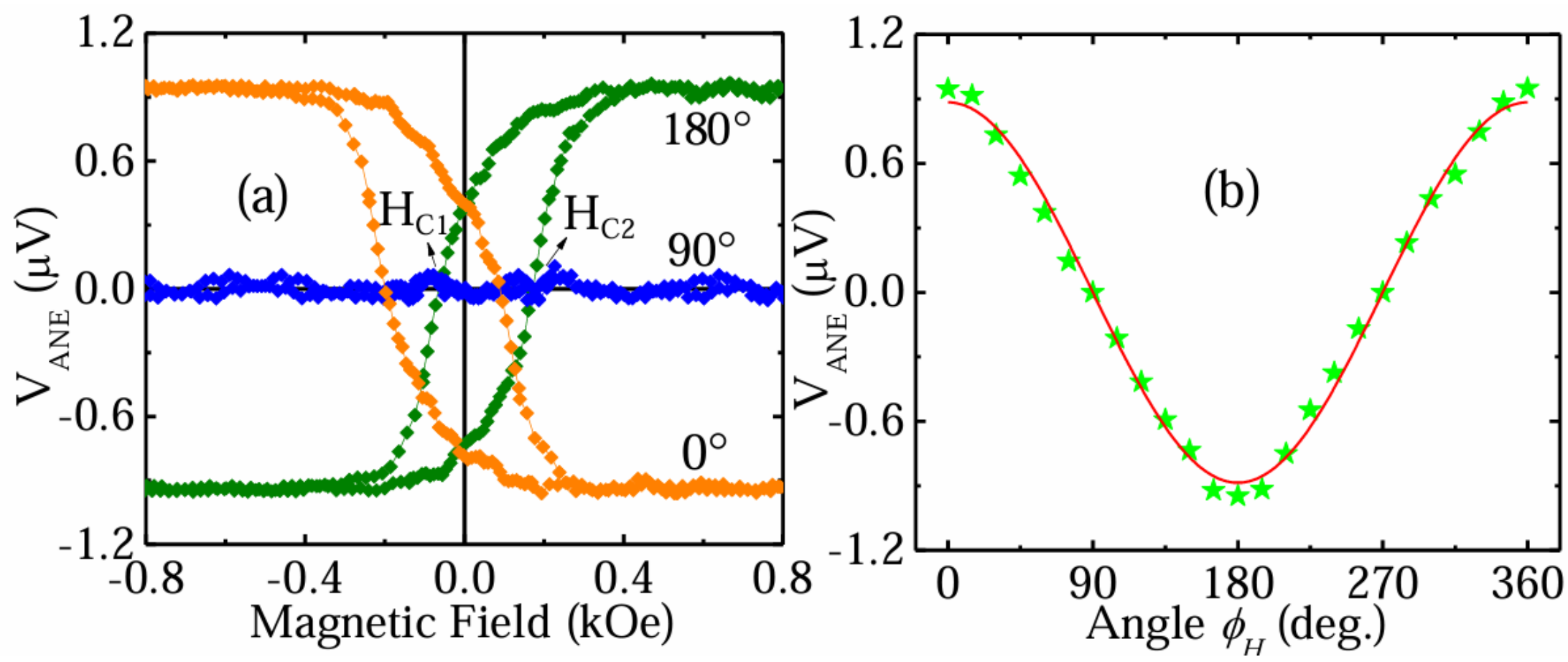

**Figure 4**: **(a)** Anomalous Nernst voltage as a function of magnetic field under $\Delta T_z=10$ K for $\phi_H=0°$, 90°, and 180°. The horizontal displacement reflects the exchange-bias field at the $IrMn_3$/Py interface. **(b)** Angular dependence of the ANE voltage at H=0.9 kOe.

Using the coordinate system of **Fig. 1**, for $H\geq H_S$ (saturation field), one obtains $\Delta V_{ANE}(\phi_H)=-\alpha_N\Delta T_z\cos\phi_H$, where $\Delta V_{ANE}=V_{ANE}-V_{0y}^{ANE}$. Since $V_{0y}^{ANE}\approx0$, we have $\Delta V_{ANE}\approx V_{ANE}$. From the fit of **Fig. 5(a)**, the anomalous Nernst coefficient is $\alpha_N=-94.8\pm0.1$ nV/K. This value is larger than that reported in Ref. [44] for planar Nernst measurements, while remaining below the ordinary anomalous Nernst coefficient $\alpha_0$ =2.7 μV/K, estimated using S=−21 μV/K [24] and $\chi\approx0.12$ [45,46]. Therefore, the obtained coefficient lies in the physically expected range for Py-based metallic films and confirms the reliability of the thermal measurements. Because **Fig. 4(b)** is measured at a fixed magnetic field in the saturation regime, the exchange shift does not appear explicitly in the angular scan. Instead, the role of $IrMn_3$ is to stabilize the fully aligned magnetization state from which the thermoelectric response is measured.

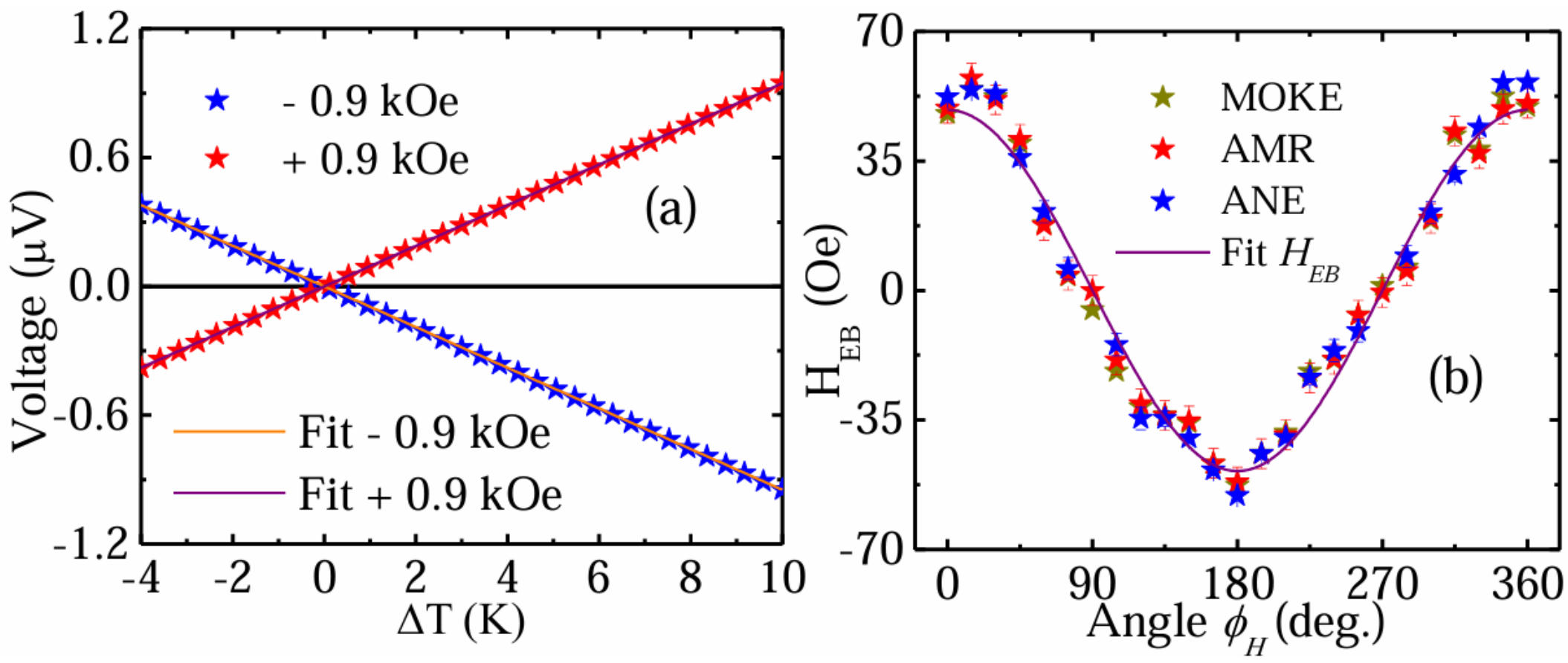

**Figure 5: (a)** Anomalous Nernst voltage as a function of the out-of-plane temperature gradient for H=±0.9 kOe, confirming linear thermoelectric scaling. **(b)** Angular dependence of the exchange-bias field extracted independently from MOKE, AMR, and ANE measurements.

As expected from the transverse symmetry of the anomalous Nernst effect, the signal vanishes when the magnetization becomes collinear with the induced electric field, making these angular positions more sensitive to experimental noise. Error bars in **Fig. 5(b)** were determined from the voltage fluctuations measured near these zero-crossing points for positive and negative field sweeps. The angular dependence of the exchange-bias field obtained from the three independent techniques is well described by $H_{EB}=H_0^{EB}\cos\phi_H$, with fitted amplitude $H_0^{EB}$=50 Oe. This value is in quantitative agreement with the MOKE and AMR determinations, demonstrating remarkable consistency across magnetic, transport, and thermoelectric observables.

Such convergence is a strong indicator of interfacial coherence in the $IrMn_3$/Py bilayer. If substantial interfacial disorder, rotatable anisotropy, or stochastic pinning dominated the system, discrepancies between the three probes would be expected. Instead, the same exchange field is recovered in all cases, confirming that the non-collinear $IrMn_3$ layer acts as a robust and reproducible symmetry-breaking element. From a broader perspective, these results reinforce the growing view that $IrMn_3$ should not be regarded merely as a conventional pinning layer. Owing to its triangular antiferromagnetic spin structure and associated anisotropic transport properties [4, 42], $IrMn_3$ constitutes an active material platform capable of simultaneously controlling magnetic states, charge transport, and spin-caloritronic conversion. The $IrMn_3$/Py heterostructure therefore emerges as a promising candidate for multifunctional antiferromagnetic spintronic devices.

**4. Conclusion**

We have demonstrated that exchange bias in $IrMn_3$/Py heterostructures constitutes a robust and reproducible interfacial symmetry-breaking field that simultaneously governs magnetic reversal, charge transport, and thermoelectric conversion. A unidirectional anisotropy of $H_{EB}$=50 Oe was independently resolved by longitudinal MOKE, anisotropic magnetoresistance (AMR), and anomalous Nernst effect (ANE) measurements, all exhibiting the same characteristic angular dependence, $H_{EB}(\phi_H)=H_0\cos\phi_H$, which confirms the common exchange-anisotropy origin of the three responses. The quantitative agreement among these complementary probes provides strong evidence of a highly coherent $IrMn_3$/Py interface, where the interfacial exchange field is not limited to shifting magnetic hysteresis loops, but is directly encoded into electrical resistance and heat-driven voltage generation. This result significantly broadens the conventional interpretation of exchange bias, establishing it as a multifunctional interfacial control parameter that can be read through magnetic, transport, and spin-caloritronic channels.

A central outcome of this work is the active role played by $IrMn_3$ as a non-collinear antiferromagnet. Unlike conventional collinear pinning materials, $IrMn_3$ possesses a triangular Mn spin structure with nearly compensated bulk magnetization, strong internal exchange interactions, and symmetry-dependent relativistic transport properties. These intrinsic characteristics enable deterministic exchange coupling while preserving the advantages of zero stray fields and magnetic robustness. In addition, $IrMn_3$ has been associated with large spin Hall responses and Berry-curvature-related phenomena, making it particularly attractive for multifunctional spintronic architectures. Our results therefore show that $IrMn_3$ should not be regarded merely as a passive exchange-bias layer. Instead, it behaves as an active antiferromagnetic component whose non-collinear spin texture can simultaneously stabilize the ferromagnetic state and influence charge and thermal transport. The observation that the same interfacial field controls MOKE, AMR, and ANE responses highlights the efficiency with which the magnetic symmetry of $IrMn_3$ is transferred to the adjacent ferromagnet.

From a technological perspective, this combination of exchange anisotropy, electrical readability, and thermoelectric functionality is highly desirable for next-generation devices. Exchange-biased $IrMn_3$/Py bilayers offer a promising route toward compact magnetic reference elements, field-free memory architectures, thermally assisted logic, spin-caloritronic sensors, and low-power multifunctional circuits. More broadly, the coexistence of magnetic stability with transport functionality makes non-collinear antiferromagnets especially appealing for scalable spintronic systems. In summary, the present work establishes the $IrMn_3$/Py heterostructure as both a model platform for studying exchange bias at ferromagnet/non-collinear-antiferromagnet interfaces and a technologically relevant materials system in which spin, charge, and heat can be coherently controlled through interfacial magnetic symmetry. These findings reinforce the emerging importance of non-collinear antiferromagnets as key materials for future spintronic and spin-caloritronic technologies.

**Acknowledgements**

This research was supported by Conselho Nacional de Desenvolvimento Científico e Tecnológico (CNPq) with Grant Number: 300631/2025-1, Coordenação de Aperfeiçoamento de Pessoal de Nível Superior (CAPES) with Grant Number: PROAP2025UFRPE, and Fundação de Amparo à Ciência e Tecnologia do Estado de Pernambuco (FACEPE) with Grant Number: APQ-1397-3.04/24.

**Contributions**

A. J., A. D., C. E., G. C., L. P., J. A., J. L., and M. C. analyzed all the experimental measures and wrote the first version of the article. And J. H. discussed and corrected the article and supervised the work.

**Conflict of interest**

The authors declare that they have no conflict of interest.

**Data availability statement**

Data will be made available on reasonable request.